# Primordial gravitational waves: Model effects on time evolution and spectrum

Sandro M. R. Micheletti[1]

Universidade Federal do Rio de Janeiro, Campus Macaé, Macaé, RJ, Brazil

**Abstract**

With the detection of gravitational waves (GW) in recent years, many papers that use simulated GW datasets to constrain dark energy models have been published. However, these works generally consider GW generated by massive astrophysical objects, such as mergers of black holes or neutron stars. In this paper, we analyse the evolution and spectrum of GW generated in the inflationary epoch, assuming a standard slow-roll single-field inflationary scenario. Three models for the background are considered: a field theory model of interacting dark energy - the Interacting Holographic Tachyonic Model, the Holographic Dark Energy Model, and the ΛCDM. The results show significant dependence on the cosmological model, especially for the spectrum, and in particular show that an interaction between dark energy and dark matter can leave a significant imprint. Therefore, future primordial gravitational waves (PGW) datasets could be very useful for constraining dark energy models, including to probe an interaction in the dark sector of the universe.

## 1) Introduction

The direct detection of GW in recent years, by the LIGO and VIRGO collaborations has opened a new observational window for Cosmology and Astrophysics [1-9]. GW were predicted by Einstein, as solutions of the General Relativity field equations [10], and can be generated by massive astrophysical objects – such as those detected by LIGO and VIRGO, generated by binary compact mergers, or they may have originated in the early universe, forming a sthocastic GW backgroung. This sthocastic background have not yet been detected, but there is some evidence for it [11,12]. PGW may have been generated in the inflationary epoch, or in the period between the end of inflation and Big Bang Nucleosynthesis (BBN), by various mechanisms as, for instance, nonperturbative excitation of fields in the reheating phase, production and merger of primordial black holes, topological defects as cosmic strings, etc. [13]. Therefore, PGW can be a probe for high energy physics, on energy scales beyond the reach of particle accelerators. The most intriguing possibility is that, in the inflationary theory, PGW were generated by quantum fluctuations of the metric itself, so that their detection would be a direct evidence of a quantum gravity phenomenon. On the other hand, PGW would also be a probe of the expansion hystory of the universe [14-24]. Therefore, there is a growing effort to detect PGW, and several experiments are underway and other will come in the coming decades.

Pulsar timing arrays (PTA) experiments are sensitive to GW at frequencies $\sim 10^{-9} - 10^{-7}$ $Hz$, ground-based interferometers are sensitive at frequencies $\sim 10 - 10^3$ $Hz$ [25]. The main signature of PGW for frequencies $f < 10^{-16}$ is the B-mode in the polarization of the Cosmic Microwave Background (CMB) radiation. The joint analysis of Planck, BICEP2/Keck Array and other data furnishes a bound $r_{0.005} < 0.032$, with 95% C. L., being $r$ the tensor to scalar ratio, adopting a pivot scale $k^* = 0.005$ $Mpc^{-1}$ [26,27]. Assuming the consistency relation $r = -8n_T$, where $n_T$ is the tensor spectral index, this corresponds to a relative spectral energy density today $\Omega_g \sim 10^{-15}$ for frequencies $f \sim 10^{-17} Hz$ . Present and forthcoming experiments aim to detect the B-mode, for instance [28-32]. Next generation CMB space missions have been proposed with the specific purpose to detect the B-mode, for instance [33-36]. Signatures of PGW also could be detected in large scale structure of the universe [37-39]. There are also proposals to detect PGW of low frequencies ($10^{-18} < f < 10^{-16}$) by their effect on gravitational lens systems, e. g. [40]. Combined analysis of Planck and BICEP2 data furnishes constraints at frequencies $\sim 10^{-20} - 10^{-16}\, Hz$ [41].

---

[1] smicheletti@macae.ufrj.br

As already said, GW are a probe to study the evolution of the Universe, including the recent period of accelerated expansion. In this sense, many works have been dedicated to study the power of future GW datasets to constrain dark energy models, showing that, in fact, these datasets improve significantly the constraints on such models, e. g., [42-49]. There are also works exploring the possibility of probing an interaction between dark energy and dark matter using future GW datasets, e.g. [50-52]. However, the vast majority of these works consider GW generated by massive astrophysical objects, such as mergers of black holes or neutron stars. Much smaller is the number of works that study the effect of dark energy models on PGW, e. g. [53,54]. In this paper, we analyze the time evolution of PGW and their relative energy density spectrum, adopting three background models: the Interacting Holographic Tachyonic Model, the Holographic Dark Energy Model and the $\Lambda CDM$, and assuming a slow-roll single-field inflation scenario. It is important to say that, to our knowledge, this is the first work to do a detailed analysis of dark energy model effects on the PGW spectrum. We found that the PGW spectrum has a significant dependence on the model and also on the parameters of a given model, including on the coupling constant in the interacting case, mainly for frequencies $10^{-19} < f < 10^{-17}$, which correspond to modes that entered the horizon in the matter era and in the more recent period of accelerated expansion.

## 2) Gravitational waves from inflation

The metric for tensor perturbations to the flat Friedmann-Robertson-Walker (FRW) universe is given by

$$ds^2 = dt^2 - a(t)^2(\delta_{ij} + h_{ij})dx^i dx^j,$$

where $a(t)$ is the scale factor, $h_{00} = h_{0i} = 0$ and $|h_{ij}| \ll 1$. The action for these tensor perturbations can be derived by expanding the Einstein-Hilbert action up to second order in transverse traceless gauge, and is given by [55,56]:

$$S_T = \frac{M_{pl}^2}{8} \int d^4x\, a^3 \left[ \dot{h}_{ij}\dot{h}_{ij} - \frac{(\vec{\nabla} h_{ij})^2}{a^2} \right]. \quad (1)$$

From a variational principle, we obtain[2]

$$\ddot{h}_{ij} + 3\frac{\dot{a}}{a}\dot{h}_{ij} - \frac{\nabla^2 h_{ij}}{a^2} = 0 \quad (2).$$

Transforming to the Fourier space, and using the conformal time $\eta$,

$$h_{ij}(\vec{x},\eta) = \int \frac{d^3\vec{k}}{(2\pi)^{3/2}} \sum_{\alpha=(+,\times)} h_{\vec{k}}^{(\alpha)}(\eta)\, e_{ij}^{(\alpha)}(\vec{k}) e^{i\vec{k}\cdot\vec{x}}, \quad (3)$$

where $e_{ij}^{(\alpha)}(\vec{k})$ is the polarization tensor, with $+,\times$ the two polarization states, we get

$$h_{\vec{k}}^{(\alpha)\prime\prime} + 2\frac{a'}{a} h_{\vec{k}}^{(\alpha)\prime} + k^2 h_{\vec{k}}^{(\alpha)} = 0 \quad (4).$$

---

[2] It is well-known that tensor perturbations of the metric do not couple to matter perturbations, unless there is anisotropic stress, which we are not considering in this work.

Defining

$$u_{\vec{k}}^{(\alpha)} \equiv a h_{\vec{k}}^{(\alpha)} \quad (5),$$

we get

$$u_{\vec{k}}^{(\alpha)''} + \left(k^2 - \frac{a''}{a}\right) u_{\vec{k}}^{(\alpha)} = 0 \quad (6).$$

Remembering that $d\eta = \frac{da}{Ha^2}$, we can rewrite $\frac{a''}{a}$ as

$$\frac{a''}{a} = 2a^2 H^2(a) + a^3 H(a) \frac{dH(a)}{da}, (7)$$

which is $\sim a^2 H^2$. So, $k^2 \ll \frac{a''}{a}$ corresponds to super-horizon modes, and it is straightforward to verify that Eq. (6) has the general solution

$$u_{\vec{k}}^{(\alpha)}(\eta) = C_1^{(\alpha)}(k) a(\eta) + C_2^{(\alpha)}(k) a(\eta) \int \frac{d\eta}{a^2} \quad (8).$$

The first term dominates in an expanding universe, so the super-horizon modes $h_{\vec{k}}^{(\alpha)}(\eta)$ remains constant, by Eq. (5). Therefore, we can write the general solution of Eq. (4) as

$$h_{\vec{k}}^{(\alpha)}(\eta) \equiv h_{\vec{k},prim}^{(\alpha)} T(\eta, \vec{k}) \quad (9)$$

where the subscript 'prim' denotes some primordial time, when the mode left the horizon, in the inflationary period. The transfer function $T(\eta, \vec{k})$ describes the evolution of the PGW mode after this enters the horizon, in times after the end of inflation. $T(\eta, \vec{k})$ is normalized such that $T(\eta, \vec{k}) \to 1$ as $k \to 0$. When $k^2 \sim \frac{a''}{a}$, the mode decays fastly and, well inside the horizon, that is, when $k^2 \gg \frac{a''}{a}$, Eq. (6) has plane wave solutions, so that the amplitudes $h_{\vec{k}}^{(\alpha)}(\eta)$ are damped as $\frac{1}{a}$. Therefore, the evolution of the amplitudes will depend on the energy content of the universe through $a$ and, while $k^2 \sim \frac{a''}{a}$, through $\frac{a''}{a}$.

The energy-momentum tensor of GW is given by [57]:

$$T_{\mu\nu}^{GW} = \frac{M_{pl}^2}{4} \langle \partial_\mu h_{ij} \partial_\nu h^{ij} \rangle.$$

The GW energy density is given by the 0-0 component of the GW energy-momentum tensor

$$\rho_g \equiv T_{00}^{GW} = \frac{M_{pl}^2}{4a^2} \langle h'_{ij}(\vec{x}, \eta) h'^{ij}(\vec{x}, \eta) \rangle. (10)$$

Defining the GW power spectrum $\Delta_h^2(k)$ as

$$\Delta_h^2(k) \equiv \frac{d\langle h_{ij} h^{ij} \rangle}{d \ln k}, (11)$$

and using Eq. (9), $\rho_g$ can be written as

$$\rho_g = \frac{M_{pl}^2}{4a^2} \int d\ln k \, \Delta_{h,prim}^2(k) [T'(\eta, \vec{k})]^2. (12)$$

The relative spectral energy density is defined as

$$\Omega_g \equiv \frac{1}{3M_{pl}^2 H^2} \frac{d\rho_g}{d\ln k}. (13)$$

Substituting Eq. (12) in (13), we get

$$\Omega_g = \frac{1}{12} \left(\frac{1}{Ha}\right)^2 \Delta_{h,prim}^2(k) [T'(\eta, \vec{k})]^2. (14)$$

The primordial power spectrum can be written as [58]:

$$\Delta_{h,prim}^2(k) = \frac{8}{M_{pl}^2}\left(\frac{H_e}{2\pi}\right)^2 \left\{1 + n_T \ln\left(\frac{k}{k_*}\right) + \frac{1}{2}\frac{dn_T}{d\ln k}\left[\ln\left(\frac{k}{k_*}\right)\right]^2\right\}, (15)$$

where $H_e$ is the Hubble parameter at the end of inflation, $n_T$ is the tensor spectral index and $k_*$ is the pivot scale.

In order to calculate $\Omega_g$ through Eq. (14), we need to solve Eq. (6), whose solutions, in general, depend on the term $\frac{a''}{a}$. In the post inflationary period, for a universe composed by dark energy, cold dark matter, baryon matter and radiation, Eq. (7) can be written as

$$\frac{a''}{a} = \frac{a^2 H^2}{2}(1 - 3\Omega_{DE}\omega - \Omega_r), (16)$$

where $\Omega_{DE}$ and $\Omega_r$ are the relative densities of dark energy and radiation, respectively, and $\omega$ is the equation of state parameter of dark energy. It is important to note that the above equation is the same, even in the case of an interaction between dark energy and dark matter. From Eq. (16) we see that the PGW evolution is affected by the dark energy model parameters at later times. Therefore, modes that entered the horizon at later times will have phase differences for different dark energy models.

The evolution of the horizon will produce differences on amplitudes for different dark energy models, as the amplitudes are damped as $\frac{1}{a}$ (remembering that we can write $a = a(\eta)$), mainly for modes that entered the horizon later. It will also genererate phase differences. In fact, as we will see below, modes that entered the horizon from the matter era onwards, that is for $k \lesssim 10 H_0$ or, equivalently, for $f \lesssim 10^{-17}$ Hz (the frequency of GW today is related to $k$ by $f = c\frac{k}{2\pi}$), are significantly affected by the dark energy model. However, it is important to note that even modes which entered the horizon earlier ($f > 10^{-17}$ Hz), suffers some influence of the dark energy model. In fact, the comoving horizon today is dependent of the dark energy model, so that $\Omega_g$ will also have some dependence on the dark energy model.

We have integrated Eq. (6) since the radiation era until the present. From Eq. (16), in the radiation era $\frac{a''}{a} = 0$, so that Eq. (6) is a plane wave equation. Imposing appropriate boundary conditions [59], we have the solution

$$u_{\vec{k}}(\eta) = a h_{\vec{k},prim} \frac{\sin(k\eta)}{k\eta} = H_0\sqrt{\Omega_{r0}} h_{\vec{k},prim} \frac{\sin(k\eta)}{k}, (17)$$

where $h_{\vec{k},prim}$ is the amplitude at some primordial time, when the mode left the horizon, in the inflationary period. Note that we have omitted the index α for polarization. In the last equality we have used that, in the radiation era, $\eta = \frac{a - a_e}{H_0\sqrt{\Omega_{r0}}} \cong \frac{a}{H_0\sqrt{\Omega_{r0}}}$, supposing $a \gg a_e$, where $a_e$ is the scale factor at the end of inflation.

The solution (17) corresponds to

$$h_{\vec{k}}(\eta) = h_{\vec{k},prim}\frac{\sin(k\eta)}{k\eta} (18).$$

Note that, for super-horizon scales, $k\eta \ll 1$, $h_{\vec{k}}(\eta) \to h_{\vec{k},prim}$ and $h'_{\vec{k}}(\eta) \to 0$, that is, the mode has constant amplitude, consistent with the super-horizon solution (8).

## 3) The background models

In this section we derive the Interacting Holographic Tachyonic Model (hereafter IHTM). This is a field theory model in which dark energy is associated with a tachyonic field, minimaly coupled with a massive fermionic field, which in turn is associated with dark matter. This model was developed and compared with observational data in [60,61]. Instead of choosing a particular form for the tachyonic potential, this is implicitly specified by imposing that tachyonic dark energy behaves like holographic dark energy. For more details, see [60,61]. As we will see below, when the coupling constant is zero, the IHTM reproduces the Holographic Dark Energy Model (hereafter HDE), developed in [62,63]. The HDE has been widely studied since it was proposed, and many variations have emerged, see for instance [64-68]. For a review on holographic models and more references, see [69]. In 3.A we present the HDE with an interaction with dark matter, and in 3.B we present a combination of this with a field theory model, in order to obtain the IHTM.

### 3.A) Interacting holographic dark energy

The density of the holographic dark energy is given by [62,63]

$$\rho_{DE} = \frac{3M_{pl}^2 c^2}{R_h^2}, \quad (19)$$

where $c$ is a free parameter and $R_h$ is the *event horizon*, given by

$$R_h(t) = a(t) \int_t^\infty \frac{dt'}{a(t')}. \quad (20)$$

In a flat FRW universe composed by dark energy and dark matter in interaction, baryonic matter and radiation, the conservation equations are

$$\dot{\rho}_{DE} + 3H\rho_{DE}(1+\omega) = Q \quad (21)$$

$$\dot{\rho}_{DM} + 3H\rho_{DM} = -Q \quad (22)$$

$$\dot{\rho}_b + 3H\rho_b = 0 \quad (23)$$

$$\dot{\rho}_r + 4H\rho_r = 0, \quad (24)$$

where dot represents derivative with respect to time and $Q$ is the interaction term. The Friedmann equation for a flat universe is

$$H^2 = \frac{1}{3M_{pl}^2}(\rho_{DE} + \rho_{DM} + \rho_b + \rho_r). \quad (25)$$

Using Eqs. (21)-(25) it is possible to rewrite Eq. (21) as

$$\dot{\Omega}_{DE} = 3H\Omega_{DE}\left[-(1-\Omega_{DE})\omega + \frac{\Omega_r}{3}\right] + \frac{Q}{3M_{pl}^2 H^2}. \quad (26)$$

On the other hand, the relative energy density of the holographic dark energy can be written as

$$\Omega_{DE} = \frac{c^2}{H^2 R_h^2}. \quad (27)$$

Deriving Eq. (19) with respect to time and using Eq. (20) and Eq. (27), we obtain

$$\dot{\rho}_{DE} = 2H\rho_{DE}\left(\frac{\sqrt{\Omega_{DE}}}{c} - 1\right). \quad (28)$$

Inserting Eq. (28) in Eq. (21), we obtain

$$\omega = -\frac{1}{3} - \frac{2\sqrt{\Omega_{DE}}}{3c} + \frac{Q}{3H\rho_{DE}}. \quad (29)$$

The interaction term $Q$ specifies the interaction between dark energy and dark matter. If $Q \equiv 0$, Eq. (29) reproduces the equation of state parameter of the HDE, presented in [62,63]. Below, we will determine $Q$ from a field theory model.

## 3.B) Interacting holographic tachyonic model

We consider the action

$$S = \int d^4x \sqrt{-g}\left\{-\frac{M_{pl}^2}{2}R - V(\varphi)\sqrt{1-\alpha\partial^\mu\varphi\partial_\mu\varphi} + \frac{i}{2}[\bar{\Psi}\gamma^\mu\nabla_\mu\Psi - \bar{\Psi}\gamma^\mu \overleftarrow{\nabla}_\mu\Psi] - (M-\beta\varphi)\bar{\Psi}\Psi + \sum_j \mathcal{L}_j\right\}, (30)$$

where $R$ is the curvature scalar, $\varphi$ is a tachyonic scalar field, which we will identify with dark energy, $\Psi$ is a fermionic field of mass $M$, which we will identify with dark matter and the last term contains the Lagrangian densities for the remaining fields. Note that there is a Yukawa-like interaction between the tachyonic and fermionic fields, with coupling constant $\beta$. If there was a coupling between the tachyonic field and baryonic matter, the corresponding coupling constant $\beta_b$ should satisfy the solar system constraint [70]

$$\beta_b \lesssim 10^{-2}. (31)$$

We assume $\beta_b \equiv 0$, which trivially satisfies the constraint given by Eq. (31). From a variational principle, we obtain

$$i\gamma^\mu\nabla_\mu\Psi - M^*\Psi = 0, (32)$$

$$i(\nabla_\mu\bar{\Psi})\gamma^\mu + M^*\bar{\Psi} = 0, (33)$$

where $M^* \equiv M - \beta\varphi$, and

$$\nabla_\mu\partial^\mu\varphi + \alpha\frac{\partial^\mu\varphi(\nabla_\mu\partial_\sigma\varphi)\partial^\sigma\varphi}{1-\alpha\partial^\mu\varphi\partial_\mu\varphi} + \frac{1}{\alpha}\frac{d\ln V(\varphi)}{d\varphi} = \frac{\beta\bar{\Psi}\Psi}{\alpha V(\varphi)}\sqrt{1-\alpha\partial^\mu\varphi\partial_\mu\varphi}. (34)$$

Eqs. (32) and (33) are the Dirac equation and its adjoint, in the case of an interaction between the Dirac field and the tachyonic field $\varphi$. For homogeneous fields and adopting the flat FRW metric, Eqs. (32) and (33) can be written as

$$\frac{d(a^3\bar{\Psi}\Psi)}{dt} = 0,$$

which is equivalent to

$$\bar{\Psi}\Psi = \bar{\Psi}_i\Psi_i\left(\frac{a_i}{a}\right)^3, (35)$$

where the the subscript "i" denotes some initial time. Eq. (34) reduces to

$$\ddot{\varphi} = -(1-\alpha\dot{\varphi}^2)\left[\frac{1}{\alpha}\frac{d\ln V(\varphi)}{d\varphi} - 3H\dot{\varphi} - \frac{\beta\bar{\Psi}\Psi}{\alpha V(\varphi)}\sqrt{1-\alpha\dot{\varphi}^2}\right]. (36)$$

From the energy–momentum tensor, we obtain.

$$\rho_{DE} = \frac{V(\varphi)}{\sqrt{1-\alpha\dot{\varphi}^2}}, (37)$$

$$P_{DE} = -V(\varphi)\sqrt{1-\alpha\dot{\varphi}^2}, (38)$$

$$\rho_{DM} = M^*\bar{\Psi}\Psi, (39)$$

$$P_{DM} = 0. (40)$$

From Eqs. (37) and (38) we have

$$\omega \equiv \frac{P_{DE}}{\rho_{DE}} = \alpha\dot{\varphi}^2 - 1. (41)$$

Deriving Eqs. (37) and (39) with respect to time and using Eqs. (35) and (36), we obtain

$$\dot{\rho}_{DE} + 3H\rho_{DE}(1 + \omega) = \beta\dot{\varphi}\bar{\Psi}_i\Psi_i\left(\frac{a_i}{a}\right)^3 \quad (42)$$

and

$$\dot{\rho}_{DM} + 3H\rho_{DM} = -\beta\dot{\varphi}\bar{\Psi}_i\Psi_i\left(\frac{a_i}{a}\right)^3, \quad (43)$$

where the dot represents derivative with respect to time. Comparing Eqs. (42) and (43) with Eqs. (21) and (22), we see that

$$Q = \beta\dot{\varphi}\bar{\Psi}_i\Psi_i\left(\frac{a_i}{a}\right)^3. \quad (44)$$

Using Eq. (39) and remembering that $\rho_{DMi}=3M_{pl}^2 H_i^2 \Omega_{DMi}$, we have

$$\bar{\Psi}_i\Psi_i = \frac{3M_{pl}^2 H_i^2 \Omega_{DMi}}{M - \beta\varphi_i}. \quad (45)$$

From Eq. (41) we have

$$\dot{\varphi} = \frac{sign[\dot{\varphi}]}{\sqrt{\alpha}}\sqrt{1+\omega}. \quad (46)$$

Substituting Eqs. (45) and (46) in Eq. (44), and defining $\phi \equiv \sqrt{\alpha}\varphi$, we have

$$Q = \delta\, sign[\dot{\phi}]3M_{Pl}^2 H_i^2 \Omega_{DMi}\sqrt{1+\omega}\left(\frac{a_i}{a}\right)^3, \quad (47)$$

where we have defined the effective coupling constant

$$\delta \equiv \frac{\frac{\beta}{\sqrt{\alpha}M}}{1 - \frac{\beta}{\sqrt{\alpha}M}\phi_i}. \quad (48)$$

$sign[\dot{\phi}]$ is in fact arbitrary, as it can be changed by redefinitions of the tachyon field, $\varphi \to -\varphi$, and of the coupling constant $\beta \to -\beta$. Moreover, as the observable is $\delta$, $\phi_i$ can be chosen arbitrarily, and we take $\phi_i \equiv 0$ [60,71]. Substituting Eq. (47) in Eq. (26) we obtain

$$\frac{d\Omega_{DE}}{dz} = \frac{3\Omega_{DE}}{1+z}\left\{(1 - \Omega_{DE})\omega - \frac{\Omega_r}{3} - \sqrt{\frac{2}{3}}\gamma\sqrt{1+\omega}\right\}, \quad (49)$$

where

$$\gamma = \frac{\delta}{\sqrt{6}}\frac{H_i^2}{H^3}\frac{\Omega_{DMi}}{\Omega_{DE}}\left(\frac{1+z}{1+z_i}\right)^3. \quad (50)$$

It is convenient to write Eq. (50) in terms of today values,

$$\gamma = \frac{1}{\sqrt{6}}\left(\frac{\delta}{H_0}\right)\frac{1}{E^3}\frac{\Omega_{DM0}}{\Omega_{DE}}\left(\frac{1+z}{1+z_0}\right)^3, \quad (50.a)$$

where $\Omega_{DM0}$ is the relative energy density of cold dark matter today and $E \equiv \frac{H}{H_0}$. Evidently, $\Omega_{DM0} = 1 - \Omega_{DE0} - \Omega_{b0} - \Omega_{r0}$ for a flat universe. Here, it is worth to comment that $\sqrt{\alpha} \sim (H_0 \times MeV)^{-1}$, so $\frac{\delta}{H_0}$ is dimensionless - see Eq. (48), such that it is more convenient to constrain $\frac{\delta}{H_0}$, as we have done in [60,61].

Combining Eqs. (47) and (29) and solving for $\omega$, we obtain

$$\omega = -\frac{1}{3} - \frac{2\sqrt{\Omega_{DE}}}{3c} + \frac{\gamma}{3}\left[\gamma + \sqrt{\gamma^2 + 4\left(1 - \frac{\sqrt{\Omega_{DE}}}{c}\right)}\right], (51)$$

or, solving for $Q$, we obtain

$$Q = \frac{1}{\sqrt{6}}\left(\frac{\delta}{H_0}\right) H_0 \rho_{DM0} \left(\frac{1+z}{1+z_0}\right)^3 \left[\gamma + \sqrt{\gamma^2 + 4\left(1 - \frac{\sqrt{\Omega_{DE}}}{c}\right)}\right]. \quad (52)$$

Note that we have expressed $Q$ in terms of $\left(\frac{\delta}{H_0}\right)$, which is dimenionless, as said above. It is interesting to comment that, for $z > 0$, $\gamma$ rapidly goes to very small values, so that $Q \cong \sqrt{\frac{2}{3}}\left(\frac{\delta}{H_0}\right) H_0 \rho_{DM}(z)$, similar to a phenomenological form of $Q$, widely used in the literature[3].

From Eq. (23) we have

$$\rho_b(z) = \rho_{b0}\left(\frac{1+z}{1+z_0}\right)^3,$$

where $\rho_{b0} = 3M_{pl}^2 H_0^2 \Omega_{b0}$. Using Eqs. (21)-(25) it is possible to rewrite Eq. (24) as

$$\frac{d\Omega_r}{dz} = -\frac{3\Omega_r}{1+z}\left\{\Omega_\phi \omega + \frac{\Omega_r}{3} - \frac{1}{3}\right\}. \quad (53)$$

The Hubble parameter can be written as

$$H = \lambda \frac{(1+z)^2}{\sqrt{\Omega_r}}, \quad (54)$$

where $\lambda \equiv \frac{\pi}{3M_{pl}}\sqrt{\frac{1+0.2271 N_{eff}}{5}} T_{CMB}^2$, $N_{eff} = 3.04$ is the effective number of relativistic degrees of freedom, $T_{CMB} = 2.725\,K$ is the CMB temperature today. Note, from Eq. (54), that $H(z=0) = \frac{\lambda}{\sqrt{\Omega_{r0}}} = H_0$.

Eq. (49) is general for interacting dark energy models. A given model is specified by $\omega$ and $\gamma$, which in the case of the IHTM are given by Eqs. (51) and (50), respectively. For a noninteracting dark energy model, $\gamma \equiv 0$, and the model is specified by $\omega$. As already mentioned before, the HDE can be obtained as a particular case of the IHTM with $\delta = 0$. For the ΛCDM, $\gamma \equiv 0$ and $\omega = -1$.

It is worth to comment that the interaction affects the evolution of dark matter even when dark energy is negligible. In fact, from Eq. (52) we get $Q \propto \delta(1+z)^3$, for $z \gg 0$, so that the time evolution of dark matter, given by Eq. (43), and consequently of the scale factor $a$ are affected by the interaction, even in the matter era, when the dark energy density is negligible. Therefore, the interaction affects the evolution of PGW and also the amplitude of the spectrum, even for modes that entered the horizon in the matter era, which correspond to frequencies $f < 10^{-17}Hz$, as we will see below.

## 4) PGW evolution and spectrum

The evolution of PGW depends on the dynamics of the components of the universe, as it is damped as $\frac{1}{a}$ and depends on $\frac{a''}{a}$, according to Eqs. (5) and (6). The expression of $\frac{a''}{a}$ in terms of the components of the universe is given by Eq. (16). The free parameters for the IHTM are $\Omega_{DE0}$, $\Omega_{b0}$, $h$, $c$ and $\delta$. The HDE can be obtained as a particular case of IHTM, with $\delta = 0$. The free parameters for the ΛCDM are $\Omega_{DE0}$, $\Omega_{b0}$ and $h$. In this work, we have fixed $h = 0.67$ and $\Omega_{b0} = 0.05$, in agreement with the constraints presented in [61] and [72]. In fact, as shown below, the PGW time evolution and, consequently, its spectrum, depends signifcantly on dark energy model parameters. The spectra were obtained from Eq. (14), assuming a single field slow-roll inflationary scenario, in which $n_T = -\frac{r}{8}$, with $r = 0.032$, which saturates the upper limit of 95% confidence level, $r_{0.005} < 0.032$, from the joint analysis of Planck and BICEP2/Keck Array data [26,27], the pivot scale $k_* = 0.05\ Mpc^{-1}$ and initial radiation temperature $T_R = 10^{16} GeV$.

---

[3] Because of the interaction, $\rho_{DM0}\left(\frac{1+z}{1+z_0}\right)^3$ is not $\rho_{DM}(z)$. However, the coupling constant is small, tipically $\frac{\delta}{H_0} \sim 10^{-1}$, so that $\rho_{DM}(z) \cong \rho_{DM0}\left(\frac{1+z}{1+z_0}\right)^3$.

It is possible to put an upper bound on $\Omega_g(f)$ from the observational data [41], through the expression

$$\Omega_g(f) = \frac{3}{128} r A_s \Omega_{r0} \left(\frac{f}{f_*}\right)^{n_T} \left[\frac{1}{2}\left(\frac{f_{eq}}{f}\right)^2 + \frac{16}{9}\right], \quad (55)$$

where $r$ is the tensor-to-scalar ratio, $A_s$ is the amplitude of the primordial power spectrum of scalar perturbations, $f_* = 1.55 \times 10^{-15} Hz\ Mpc \times k^*$ is the pivot frequency, $n_T$ is the tensor spectral index and $f_{eq} = \frac{\sqrt{2} c H_0 \Omega_{m0}}{2\pi \sqrt{\Omega_{r0}}}$ is the corresponding frequency of a mode that enters the horizon at matter-radiation equality. We have adopted $A_s = 2.101 \times 10^{-9}$ [72] and, as already said above, $k^* = 0.05 Mpc^{-1}$, $r = 0.032$ and $n_T = -\frac{r}{8}$.

Fig. 1 shows the evolution of PGW with respect to conformal time (normalized to present value), for the IHTM, with coupling constant $\frac{\delta}{H_0} = -0.2$, for the HDE ($\delta = 0$), both with c = 1.0, and for the ΛCDM. For the three models, $\Omega_{DE0} = 0.70$. Modes with $f = 10^{-18}\ Hz, f = 10^{-17}\ Hz$ and $f = 10^{-16}\ Hz$ are shown. While the mode remains larger than the comoving horizon, $k > aH$, the amplitude remains constant, as mentioned in Section 2. After the mode enters the horizon, this rapidly decays and, well inside the horizon, when $k \ll aH$, this oscillates damped as $1/a$. Therefore, this damping depends on the dark energy model, so that the amplitudes of PGW are different for different models. On the other hand, differences on the evolution of the horizon, together with the model dependence of $\frac{a''}{a}$, generate also phase differences. Such differences will reflects itself on spectral relative energy densities.

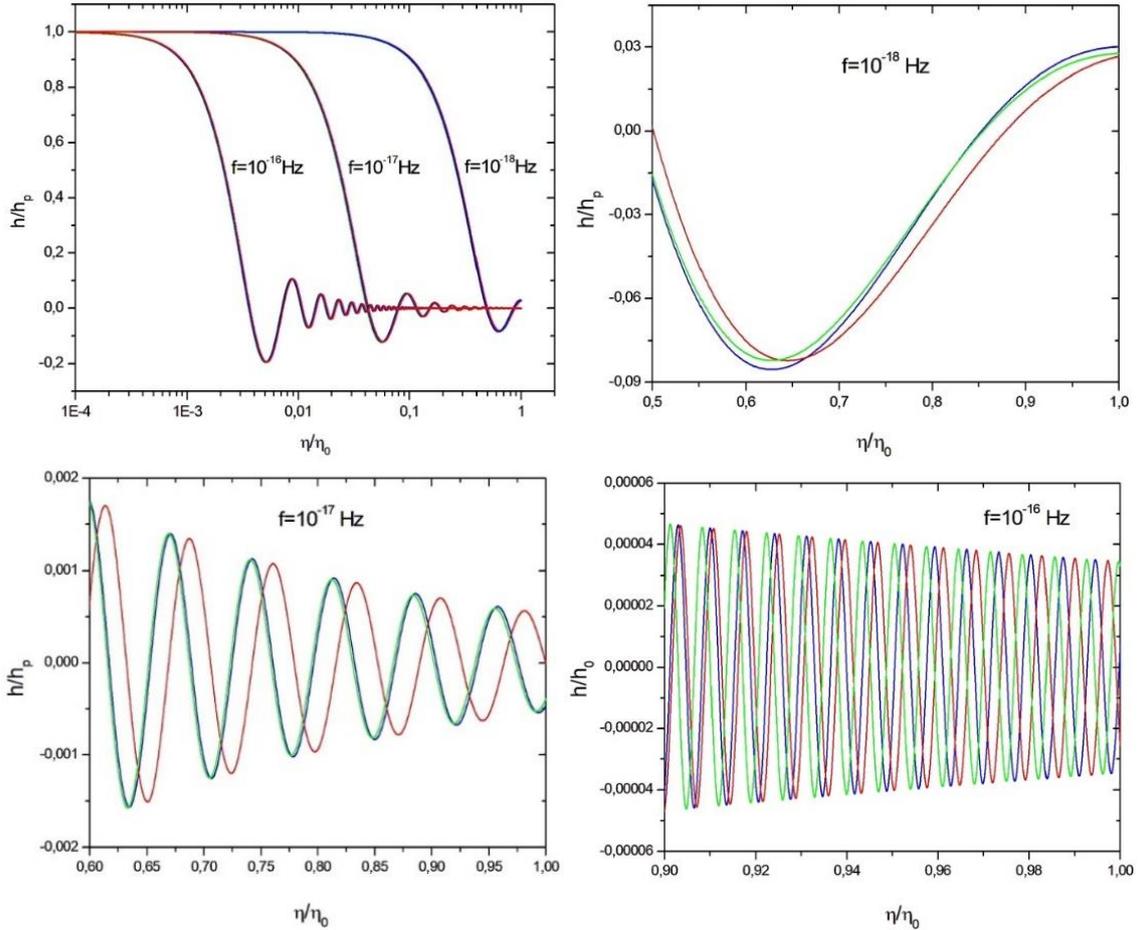

**Fig. 1**, *upper left panel: evolution of PGW with respect to conformal time, for the IHTM, with $\delta/H_0 = -0.2$ (green), HDE (red), both with $c = 1.0$, and for ΛCDM (blue). For the three models, $\Omega_{DE0} = 0.70$. Frequencies $f = 10^{-18}$ Hz, $f = 10^{-17}$ Hz and $f = 10^{-16}$ Hz are shown. Upper right panel, lower left and right panels: same as the upper left panel, but separating the frequencies $f = 10^{-18}$ Hz, $f = 10^{-17}$ Hz and $f = 10^{-16}$ Hz and enlarging in more recent times to better show the amplitude and phase differences between the three models.*

Fig. 2 shows the relative spectral energy density today, for the IHTM, HDE and ΛCDM. Again, $\Omega_{DE0} = 0.70$ for the three models, $\frac{\delta}{H_0} = -0.2$ for the IHTM and c = 1.0 for both, IHTM and HDE. The upper bound from Eq. (55) also is shown. As mentioned above, the differences in amplitude and phase in the propagation of PGW reflects itself in the relative spectral energy density. The amplitudes of the spectra have differences up to about 15% - note that in figure 2 the scale is logarithmic. The amplitude and phase differences combined can produce even more significant differences in $\Omega_g$. The points marked by diamonds (IHTM), circles (HDE) and triangles (ΛCDM) emphasizes the differences in $\Omega_g$, for some values of the frequency $f$. Note that for some values of $f$ the differences in $\Omega_g$ between the models are more than one order of magnitude.

The differences in amplitude becomes more pronounced for frequencies $f \lesssim 10^{-17}$ Hz, because such frequencies corresponds to modes that entered the horizon later, and therefore suffered less damping, so that differences in the damping due to the dark energy model are more important for these frequencies. But, in fact, some difference in amplitudes, as far as phase differences appear in the whole spectrum, even for modes that entered the horizon earlier, because the comoving horizon today is dependent of the dark energy model. Additionally, it is interesting to comment that, as modes that entered the horizon later suffer less damping, the spectrum have larger amplitude for lower frequencies.

The amplitude of the spectrum depends on the amount of dark energy: more (minus) dark energy implies in smaller (larger) amplitudes. Moreover, this depends also on the dynamics of dark energy: the three models have the same amount of dark energy today, but have different amplitudes. The detailed dynamics of dark energy depends on the equation of state parameter of dark energy ω. For the ΛCDM, $\omega = -1$; for the HDE, ω depends on parameters $\Omega_{DE0}$ and c and for the IHTM, also depends on δ, according to Eq. (51) (remembering that $h$ and $\Omega_{b0}$ are fixed, as already said above). Note that for frequencies $f \sim 10^{-19}$ Hz the spectrum is the same for the three models, which was expected, as these frequencies correspond to super-horizon modes, which have constant amplitudes.

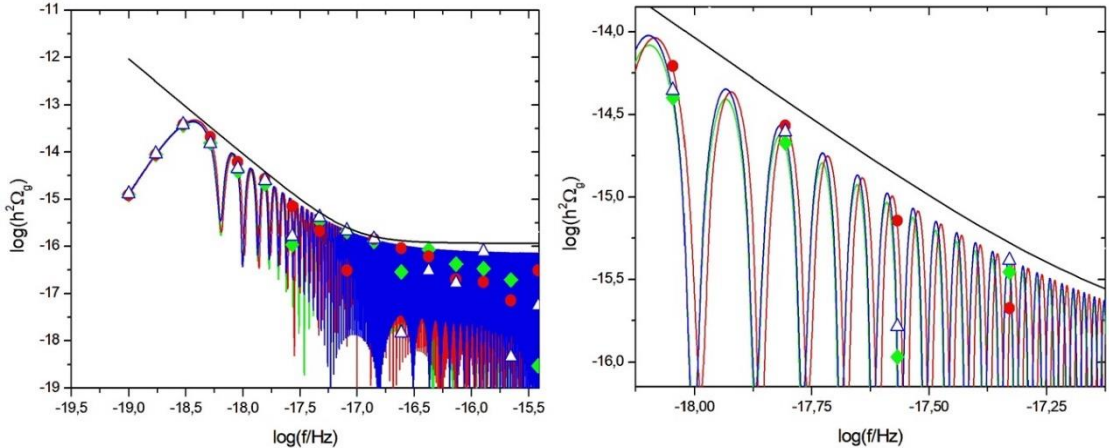

**Fig. 2**, *left panel: relative spectral energy density today of PGW, for the IHTM, with $\delta/H_0 = -0.2$ (green line and diamonds), for the HDE (red line and circles), both with $c = 1$, and for the ΛCDM (blue line and triangles). The observational bound given by Eq. (55) also is shown (black line). For the three models, $\Omega_{DE0} = 0.70$. The diamonds, circles and triangles mark $\Omega_g$ for some values of $f$, to evidence the differences in $\Omega_g$ for the three models. The right panel shows an enlargement of the left panel for $f \lesssim 10^{-17}$ Hz.*

Fig. 3 shows the effects on the evolution of PGW of varying the coupling constant δ in the IHTM, keeping fixed $c = 1$ and $\Omega_{DE0} = 0.70$. We chose just negative values of δ, as this corresponds to dark energy decaying into dark matter, alleviating the coincidence problem. Moreover, the values chose are consistent with the observational constraints presented in [61] – except the value $\frac{\delta}{H_0} = -0.7$, which was included here just as an example. For more negative values of δ, the amplitudes decrease in more recent times. This is because more negative values of δ imply in more negative ω and, therefore, in larger sizes of the comoving horizon, that is, in more expansion. There are also phase differences, because the model dependence of comoving horizon and of $\frac{a''}{a}$.

Fig. 4 shows the corresponding spectral relative energy densities. The amplitudes differ up to about 30%. Note that the coupling constant affects the amplitude of the spectrum, even in frequencies that correspond to modes that entered the horizon in the matter era, because the evolution of the scale factor $a$ is affected by the interaction even in the matter era, as already said at the end of section 3.B. Again, the combination of amplitude and phase differences can lead to differences in $\Omega_g$ of more than one order of magnitude, for some values of $f$, as can be seen by looking at the points marked by triangles. Figures 3 and 4 show that the interaction has significant effects on the evolution and spectrum of PGW.

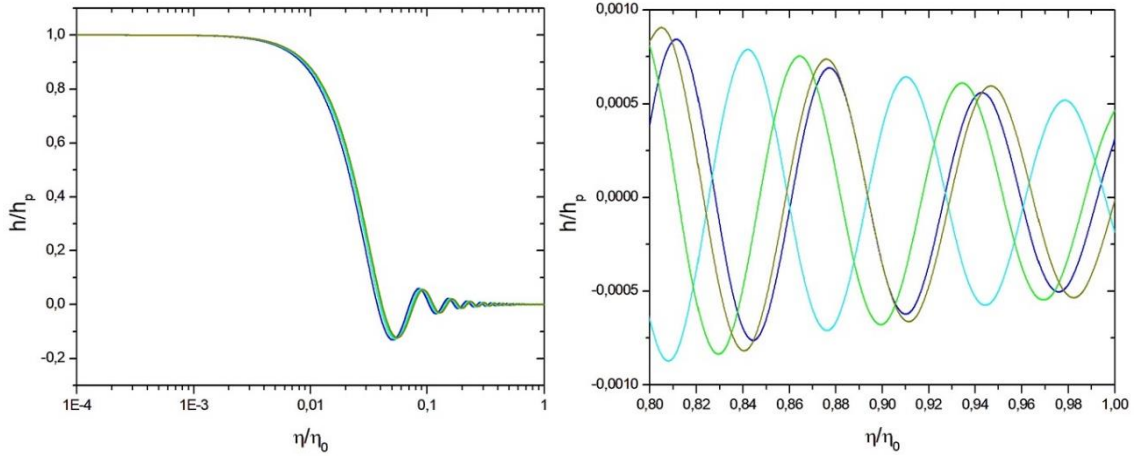

**Fig. 3**, *left panel: evolution of PGW with respect to conformal time, for IHTM, with c=1, $\Omega_{DE0} = 0.70$ and $\delta/H_0 = -0.1$, (dark yellow), $\delta/H_0 = -0.2$ (green), $\delta/H_0 = -0.4$ (cyan) and $\delta/H_0 = -0.7$ (blue). The right panel is the same as the left one, but enlarging the more recent period, in order to emphasize the amplitude and phase differences.*

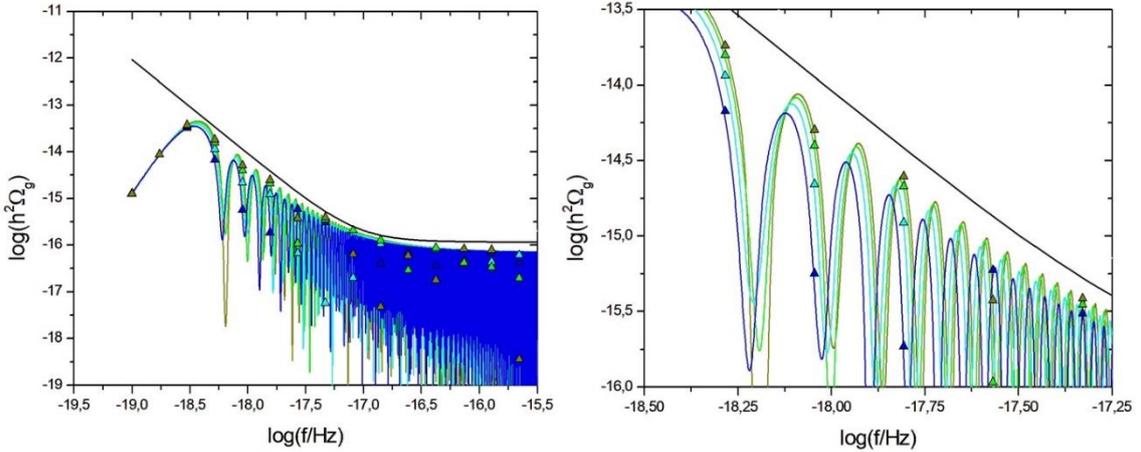

**Fig. 4**, *left panel: relative spectral energy density today of PGW, for the IHTM, with c=1, $\Omega_{DE0} = 0.70$ and $\delta/H_0 = -0.1$ (dark yellow), $\delta/H_0 = -0.2$ (green), $\delta/H_0 = -0.4$ (cyan) and $\delta/H_0 = -0.7$ (blue). The solid black line represents the observational bound given by Eq. (55). The triangles mark the values of $\Omega_g$ for some values of the frequency f. The right panel shows an enlargement of the left one for $f \lesssim 10^{-17} Hz$.*

Fig. 5 shows the effect on the PGW evolution of varying $\Omega_{DE0}$ in the IHTM, keeping $c = 1$ and $\frac{\delta}{H_0} = -0.2$ fixed. Note that $\Omega_{DE0}$ values was varied such that the relative matter density $\Omega_{m0}$ agrees with recent observational constraints [9,73-75] (For more detais, see the webpage [76]). Moreover, the $\Omega_{DE0}$ values are consistent with the observational constraints presented in [60,61]. For decreasing values of $\Omega_{DE0}$, the amplitude increases. This is because decreasing amounts of dark energy implies in decreasing sizes of the comoving horizon, that is, in less expansion, so that the waves are less damped. There are also phase differences, because the dependence of comoving horizon and $\frac{a''}{a}$ on $\Omega_{DE0}$.

Fig. 6 shows the corresponding spectral relative energy densities. Black lines correspond to observational limits of Eq. (55). Now, the differences in amplitudes are more pronounced than in Figs. 2 and 4, reaching around 50%. Triangles mark the values of $\Omega_g$ for some values of the frequency $f$. Again, the differences in $\Omega_g$ are more than one order of magnitude, in some cases.

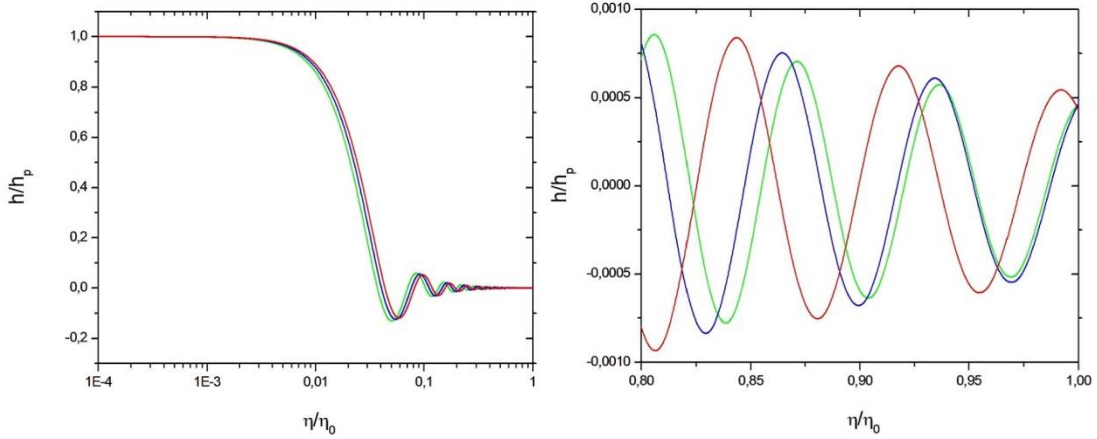

**Fig. 5**, *left panel: evolution of PGW with respect to conformal time, for the IHTM, for $f = 10^{-17}$ Hz, with $c = 1$ and $\delta/H_0 = -0.2$ and $\Omega_{DE0} = 0.65$ (red), $\Omega_{DE0} = 0.70$ (blue) and $\Omega_{DE0} = 0.75$ (green). Right panel: same as the left one, but enlarging the more recent period, in order to emphasize the amplitude and phase differences.*

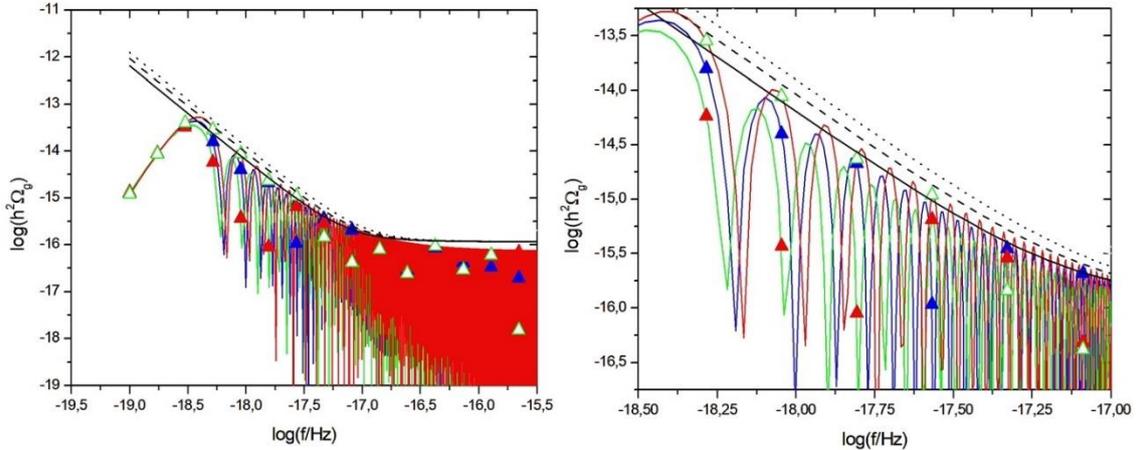

**Fig. 6**, *left panel: relative spectral energy density today of PGW, for the IHTM, with with $c = 1.0$, $\frac{\delta}{H_0} = -0.2$ and $\Omega_{DE0} = 0.65$ (red), $\Omega_{DE0} = 0.70$ (blue), $\Omega_{DE0} = 0.75$ (green), with the corresponding observational bounds given by Eq. (55) represented by dotted ($\Omega_{DE0} = 0.65$), dashed ($\Omega_{DE0} = 0.70$) and solid ($\Omega_{DE0} = 0.75$) black lines. Right panel: same as the left one, enlarged for $f \lesssim 10^{-17}$ Hz.*

Figs. 7 and 8 are analogous to Figs. 5 and 6, but now for the HDE, with $c = 1$. Again, varying $\Omega_{DE0}$ produces very pronounced differences in the spectral energy density $\Omega_g$.

The same comments applies for the $\Lambda$CDM, shown in Figs. 9 and 10. The case $\Omega_{DE0} = 0$ (CDM model) also is shown, for comparison. Note also that the upper bound of Eq. (55) is always respected (the upper bound corresponding to the CDM model is the dotted-dashed black line). The damping effect due to dark energy becomes more evident.

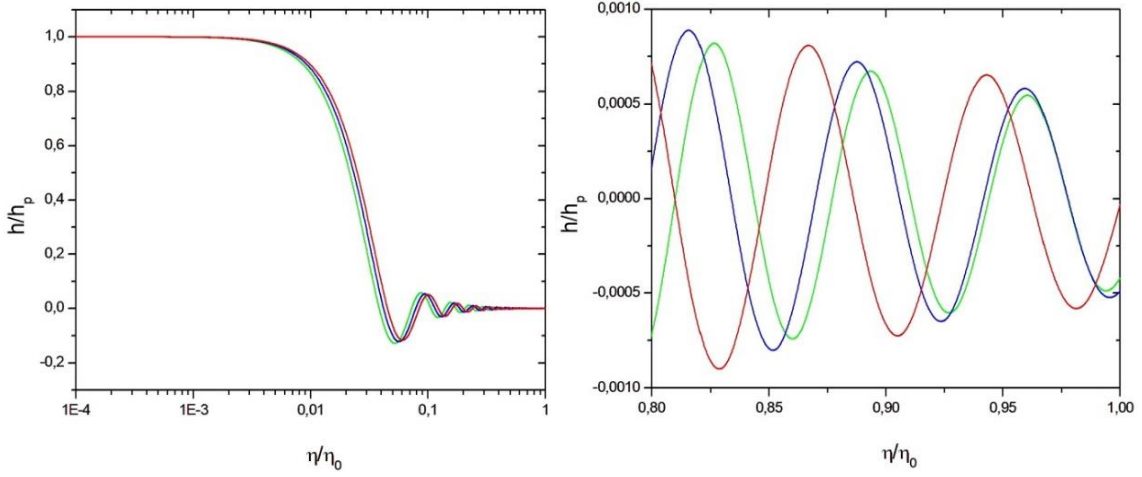

**Fig. 7**, *left panel: evolution of PGW with respect to conformal time, for the HDE, for $f = 10^{-17}$ Hz, with $c = 1$, $\Omega_{DE0} = 0.75$ (green), $\Omega_{DE0} = 0.70$ (blue) and $\Omega_{DE0} = 0.65$ (red). Right panel: same as left one, but enlarging the more recent period, in order to emphasize the amplitude and phase differences.*

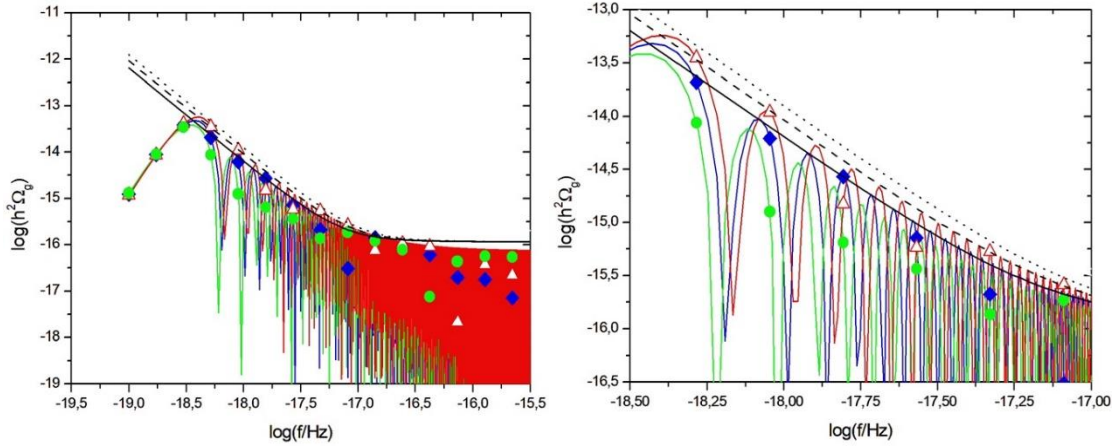

**Fig. 8**, *left panel: : relative spectral energy density today of PGW, for the HDE, with c=1 and $\Omega_{DE0} = 0.65$ (red), $\Omega_{DE0} = 0.70$ (blue) and $\Omega_{DE0} = 0.75$ (green), with the corresponding observational bounds given by Eq. (55) represented by dotted ($\Omega_{DE0} = 0.65$), dashed ($\Omega_{DE0} = 0.70$) and solid ($\Omega_{DE0} = 0.75$) black lines. Right panel: same as the left one, enlarged for $f \lesssim 10^{-17}$ Hz.*

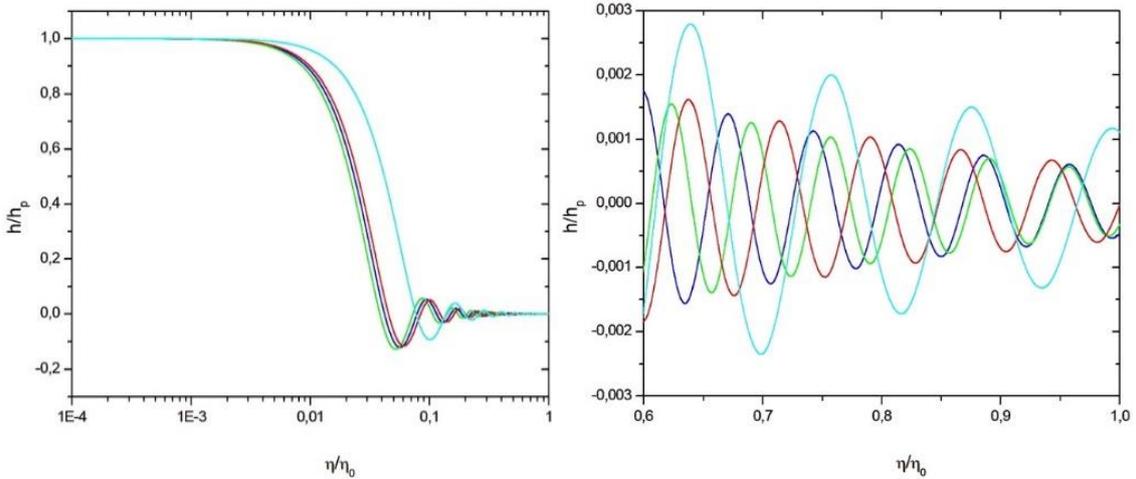

**Fig. 9**, *left panel: evolution of PGW with respect to conformal time, for $f = 10^{-17}$ Hz and ΛCDM model, with $\Omega_{DE0} = 0$ (cyan) (CDM model), $\Omega_{DE0} = 0.65$ (red), $\Omega_{DE0} = 0.70$ (blue), $\Omega_{DE0} = 0.75$ (green). Right panel: same as the left one, but enlarging the more recent period, in order to emphasize the amplitude and phase differences.*

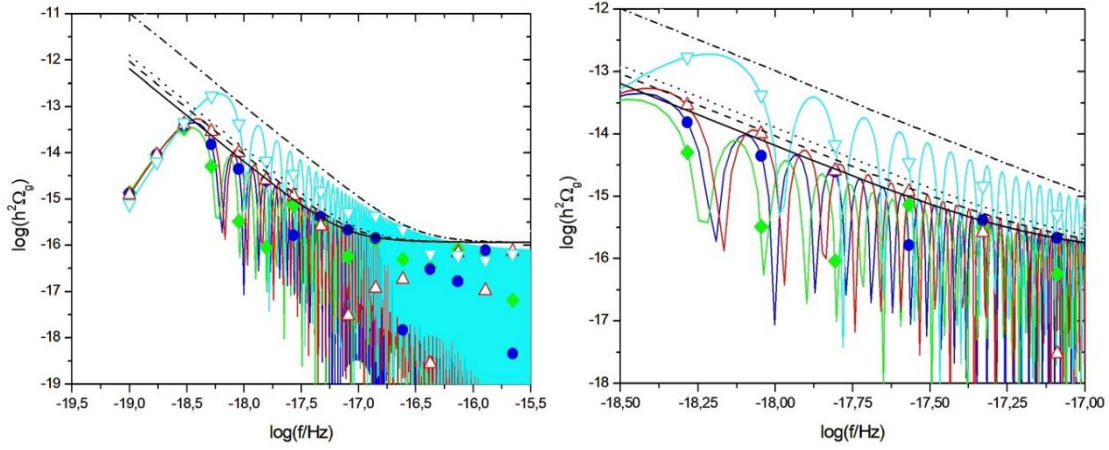

**Fig. 10**, *left panel: relative spectral energy density today of PGW, for the ΛCDM, with $\Omega_{DE0} = 0$ (cyan), $\Omega_{DE0} = 0.65$ (red), $\Omega_{DE0} = 0.70$ (blue), $\Omega_{DE0} = 0.75$ (green), with the corresponding observational bounds given by Eq. (55) represented by dotted-dashed ($\Omega_{DE0} = 0$), dotted ($\Omega_{DE0} = 0.65$), dashed ($\Omega_{DE0} = 0.70$) and solid ($\Omega_{DE0} = 0.75$) black lines. Right panel: same as the left one, enlarged for $f \lesssim 10^{-17}$ Hz.*

Fig. 11 shows the effect of varying $c$ on the evolution of PGW, in the IHTM, maintaining $\Omega_{DE0}$ fixed. The values of $c$ were varied in agreement with the observational constranits presented in [60,61], except $c = 2.0$, which was considered just as an example. Increasing values of $c$ corresponds to decreasing amplitudes. This can be understood as follows. From Eq. (16), $\frac{a''}{a} = \frac{a^2 H^2}{2}(1 + 3\Omega_{DE}|\omega|)$ from the matter era onwards. As $|\omega|$ decreases as c increases – see Eq. (51), so $a''$ also decreases as c increases, so that $a$ is larger in the past for larger values of $c$, in order to reach the same value today ($a_0 = 1$). Therefore, larger values of $c$ corresponds to smaller amplitudes. As before, the dependence of the comoving horizon and $\frac{a''}{a}$ on $c$ generate also phase differences.

In the spectral relative energy densities, shown in Fig. 12, the amplitude differences are less than 10%. However, these amplitude differences combined with phase differences, can produce very significant differences in the $\Omega_g$, for fixed values of frequency $f$, as evidenced by the points marked by triangles in figure 12.

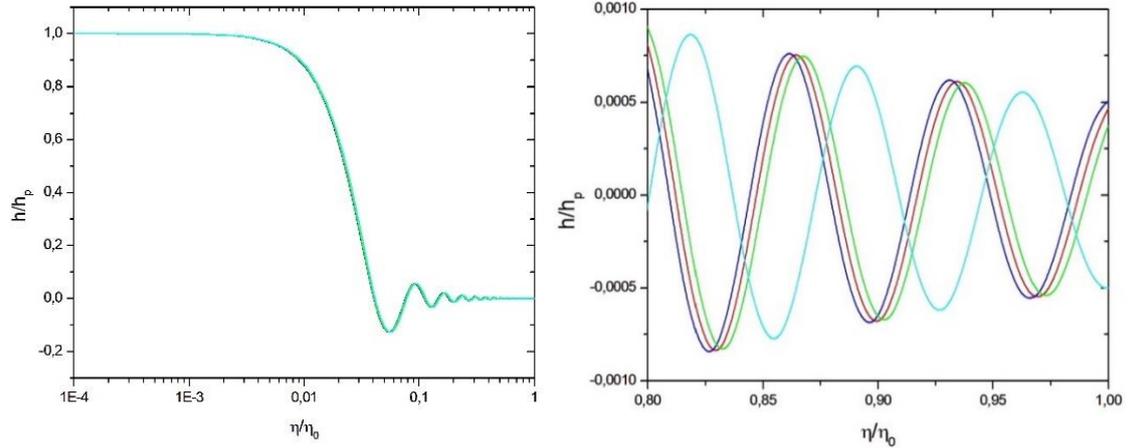

**Fig. 11**, *left panel: evolution of PGW with respect to conformal time, for $f = 10^{-17}$ Hz, for the IHTM with $\delta/H_0 = -0.2$, and c=0.9 (blue), c=1.0 (red), c=1.1 (green) and c=2.0 (cyan). Right panel: same as the left one, but enlarging the more recent period, in order to emphasize the amplitude and phase differences.*

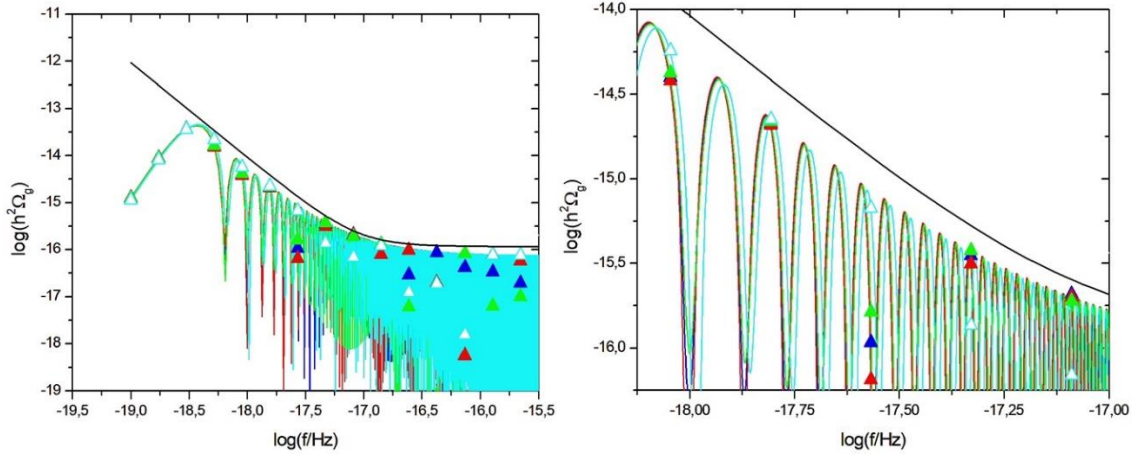

**Fig. 12**, *left panel: relative spectral energy density today of PGW, for the IHTM, with $\Omega_{DE0} = 0.7$, $\delta/H_0 = -0.2$, and c=0.9 (blue), c=1.0 (red), c=1.1 (green) and c=2.0 (cyan). The solid black line represents the observational bound given by Eq. (55). Right panel: enlargement for $f \lesssim 10^{-17}$ Hz.*

Figs. 13 and 14 are analogous to Figs. 11 and 12, but now show the HDE. The comments are the same as for Figs. 11 and 12.

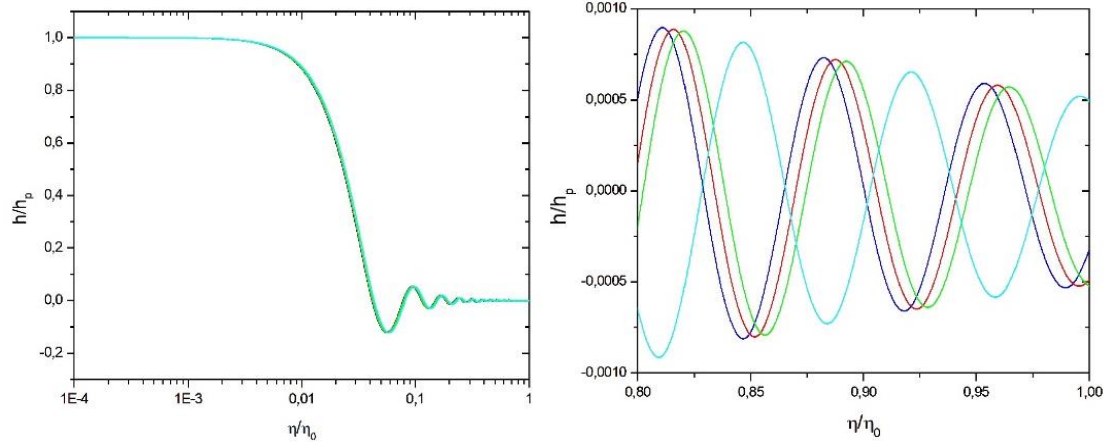

**Fig. 13**, *left panel: evolution of PGW with respect to conformal time, for $f = 10^{-17}$ Hz and for the HDE, with c=0.9 (blue), c=1.0 (red), c=1.1 (green) and c=2.0 (cyan). Right panel: same as the left one, but enlarging the more recent period, in order to emphasize the amplitude and phase differences.*

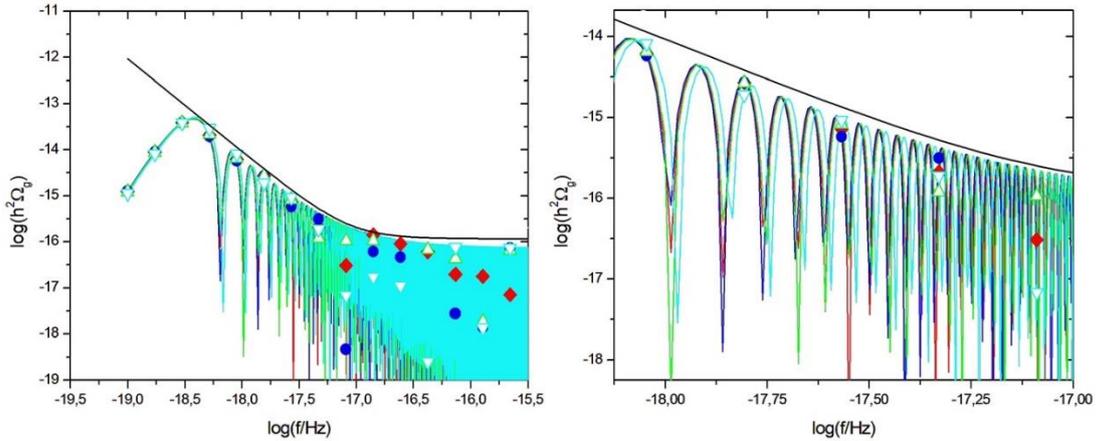

**Fig. 14**, *left panel: relative spectral energy density today of PGW, for the HDE, with $\Omega_{DE0} = 0.70$ and c=0.9 (blue) c=1 (red), c=1.1 (green) and c=2.0 (cyan). The solid black line represents the observational bound given by Eq. (55). Right panel: same as the left one, enlarged for $f \lesssim 10^{-17}$ Hz.*

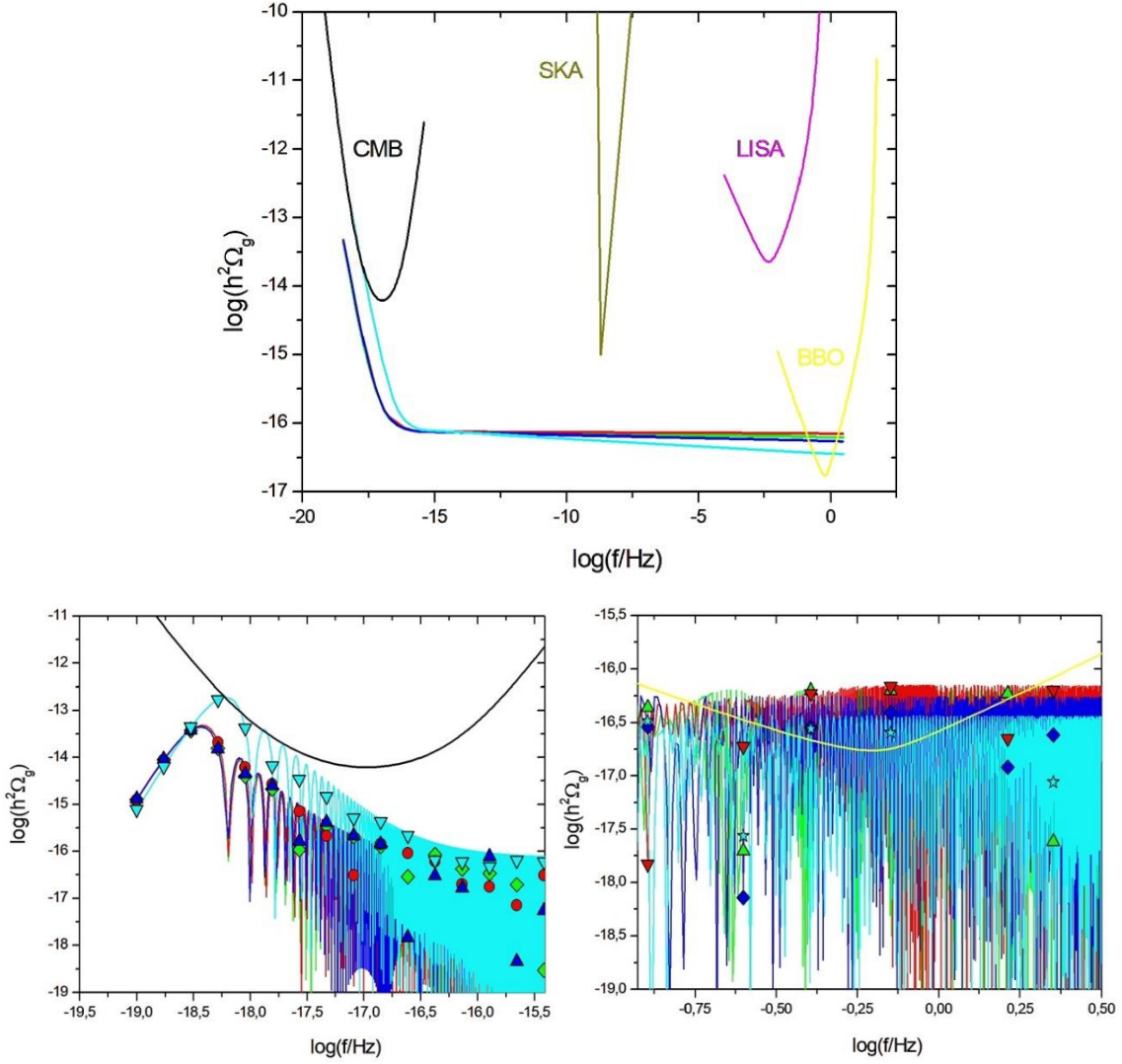

**Fig. 15**, *top panel: 95% upper limits of the CMB, SKA, LISA and BBO experiments; together with envelope curves (defined by the peaks of the relative spectral energy density today of PGW) for the IHTM, with $\delta/H_0 = -0.2$ (green curve), the HDE (red), both with c=1, the $\Lambda$CDM (blue), with $\Omega_{DE0} = 0.70$ for the three models, and the CDM (cyan); the bottom left panel shows the spectra for the same models for low frequencies, the solid black line being the 95% upper bound from CMB; the bottom right panel shows the same as the left one, but for high frequencies, the solid yellow curve being the 95% upper bound from the BBO experiment. The triangles, inverted triangles, diamonds and stars mark values of $\Omega_g$ for specific values of f, for the IHTM, HDE, $\Lambda$CDM and CDM, respectively.*

Fig. 15 shows the sensivity curves (95% upper bounds) of some forthcoming experiments and comparisons with theoretical curves. For low frequencies, $f < 10^{-16} Hz$, the observational bound on $\Omega_g$ is furnished by an analisys of the polarization of the CMB – for more details, see (13) and (41). The upper bounds of the SKA [77], LISA [78] and BBO [79], [80], [81] experiments are also shown. The 95% upper bound curve of the SKA has a minimum of $h^2\Omega_g \sim 10^{-15}$ in $f \sim 10^{-9}$, and the theoretical spectra have $h^2\Omega_g \sim 10^{-16}$ for this frequency range, so that the SKA experiment may furnish some constraint on PGW spectrum in this frequency range. Anyway, we have only upper bounds, because PGW were not detected yet. Probably, after detection, the precision will improve, and the future PGW data could furnish constraints on dark energy models. In fact, in the case of GW generated by astrophysical objects (as mergers of black holes or neutron stars), forecastings show that future GW datasets can furnish stringent constraints on dark energy models, e.g. [50-52]. There are others forthcoming experiments, as DECIGO [82], [83], MAGIS-100 [84,85], advanced LIGO+Virgo [86,87], ET [88] and CE [89]. The 95% upper bound curves of all these experiments are shown, e. g., in Fig 2 of [90]. It is interesting to mention that forecastings for the CMB-S4 data, a next-generation ground-based CMB experiment, indicate tighter constraints on PGW [91].

In the high frequencies, there is an intersection between the 95% upper bound of the BBO experiment and the theoretical spectra. Although PGW in this frequency range have entered the horizon in the radiation era, there are effects of the dark energy model on the spectra shown in the figure because, as already explained above, the comoving horizon today is model dependent. Therefore, this intersection between the BBO sensitivity curve and the theoretical PGW spectra is very promising in the sense of using future PGW datasets to constrain dark energy models.

Additionally, it is interesting to emphasize that, even in this frequency range, the interaction can leave a significant effect, as evidenced by the triangles and inverted triangles in the bottom right panel of Fig. 15, which show $\Omega_g$ for some specific values of frequency $f$, for the IHTM and the HDE, respectively. For some values of $f$, the difference in $\Omega_g$ can be more than one order of magnitude.

# 5) Conclusions

In this work we have studied the time evolution of PGW, assuming a standard slow-roll single-field inflationary scenario, adopting three models for the background: the IHTM - a field theory model of interacting dark energy, the HDE and the $\Lambda$CDM, for frequencies $10^{-18} Hz$, $10^{-17} Hz$ and $10^{-16} Hz$. Furthermore, the PGW spectral energy density today was analyzed, in the range $10^{-19} Hz \lesssim f \lesssim 10^{-15} Hz$, for the three models. There are notable differences in the results for the three models. There are also significant differences for a specific model, when the model parameters are varied.

The evolution of PGW for modes that entered the horizon from matter era onwards, what corresponds to frequencies $f \lesssim 10^{-17} Hz$, are significantly affected by the dark energy density and its dynamics, so that the amplitude of the spectrum today for these frequencies is model dependent. It is important to note, however, that even for $f > 10^{-17} Hz$ the amplitude of the spectrum suffers some influence of the model, because the size of comoving horizon today is dependent of the dark energy model. This model dependence produces also phase differences in the whole spectrum. The model dependece of the term $\frac{a''}{a}$ in Eq. (6) also produces phase differences for frequencies in the whole spectrum. In fact, the crossing of the sensitivity curve of the BBO experiment with the spectra is very promising, in the sense of using future PGW datasets to constrain dark energy models, as it shows that these model effects are detectable.

The differences in spectra can be very significant. The amplitudes of the spectra can differ by up to 50% and, for some values of $f$, due to the combination of amplitude and phase differences, $\Omega_g$ can differ by more than one order of magnitude between different models, or even for the same model, when its parameters are varied – remembering the important point that the model parameters were varied within current observational constraints. It is interesting to note that these significant differences also appeared when the coupling constant in the IHTM was varied. The fact already explained below Eq. (52), that the interaction term in the IHTM is very similar to one widely used in the literature makes this result even more interesting.

It is important to note, however, that there is also a dependence on the inflationary scenario assumed, which in principle could generate degeneracies between inflation and dark energy models. But in this case, such degeneracies could be removed by combination with other astrophysical datasets, and the PGW data would still provide additional constraints on dark energy models.

Therefore, future PGW datasets could be very useful for constraining dark energy models, including to probe an interaction in the dark sector of the universe, especially if these datasets present the relative energy density of PGW for well-defined values of the frequency.